\documentclass[superscriptaddress,prx,nofootinbib,longbibliography,twocolumn]{revtex4-1}

\usepackage{graphicx}
\usepackage{amsmath}
\usepackage{amssymb}
\usepackage{amsthm}
\usepackage{amsfonts}
\usepackage{verbatim}
\usepackage[caption=false]{subfig}
\usepackage[colorlinks,breaklinks=true]{hyperref}

\newcommand{\eq}[1]{Eq.~\hyperref[eq:#1]{(\ref*{eq:#1})}}
\renewcommand{\sec}[1]{\hyperref[sec:#1]{Section~\ref*{sec:#1}}}
\newcommand{\app}[1]{\hyperref[app:#1]{Appendix~\ref*{app:#1}}}
\newcommand{\tab}[1]{\hyperref[tab:#1]{Table~\ref*{tab:#1}}}
\newcommand{\fig}[1]{\hyperref[fig:#1]{Figure~\ref*{fig:#1}}}
\newcommand{\figa}[2]{\hyperref[fig:#1]{Figure~\ref*{fig:#1}#2}}
\newcommand{\figx}[2]{\hyperref[fig:#1]{Figure~\ref*{fig:#1}(#2)}}
\newcommand{\thm}[1]{\hyperref[thm:#1]{Theorem~\ref*{thm:#1}}}
\newcommand{\lem}[1]{\hyperref[lem:#1]{Lemma~\ref*{lem:#1}}}
\newcommand{\cor}[1]{\hyperref[cor:#1]{Corollary~\ref*{cor:#1}}}
\newcommand{\defn}[1]{\hyperref[def:#1]{Definition~\ref*{def:#1}}}
\newcommand{\alg}[1]{\hyperref[alg:#1]{Algorithm~\ref*{alg:#1}}}
\newcommand{\cubes}{n}
\newcommand{\cub}{\mu}
\newcommand{\set}{G}

\def\bra#1{\mathinner{\langle{#1}|}}
\def\ket#1{\mathinner{|{#1}\rangle}}
\newcommand{\braket}[2]{\langle #1|#2\rangle}
\newcommand{\proj}[1]{\ket{#1}\!\!\bra{#1}}

\newcommand{\be}{\begin{equation}}
\newcommand{\ee}{\end{equation}}
\newcommand{\ba}{\begin{eqnarray}}
\newcommand{\ea}{\end{eqnarray}}

\input{Qcircuit}

\begin{document}

\title{Quantum Simulation of Chemistry with Sublinear Scaling in Basis Size}

\date{\today}
\author{Ryan Babbush}
\email[Corresponding author: ]{ryanbabbush@gmail.com}
\affiliation{Google Research, Venice, CA 90291, United States}
\author{Dominic W. Berry}
\affiliation{Department of Physics and Astronomy, Macquarie University, Sydney, NSW 2109, Australia}
\author{Jarrod R. McClean}
\affiliation{Google Research, Venice, CA 90291, United States}
\author{Hartmut Neven}
\affiliation{Google Research, Venice, CA 90291, United States}

\begin{abstract}
We present a quantum algorithm for simulating quantum chemistry with gate complexity $\widetilde{\cal O}(N^{1/3} \eta^{8/3})$ where $\eta$ is the number of electrons and $N$ is the number of plane wave orbitals. In comparison, the most efficient prior algorithms for simulating electronic structure using plane waves (which are at least as efficient as algorithms using any other basis) have complexity $\widetilde{\cal O}(N^{8/3} /\eta^{2/3})$. We achieve our scaling in first quantization by performing simulation in the rotating frame of the kinetic operator using interaction picture techniques. Our algorithm is far more efficient than all prior approaches when $N \gg \eta$, as is needed to suppress discretization error when representing molecules in the plane wave basis, or when simulating without the Born-Oppenheimer approximation.
 \end{abstract}

\maketitle

\section{Introduction}
\vspace{-.1cm}

The quantum simulation of quantum chemistry is one of the most anticipated applications of both near-term and fault-tolerant quantum computing. The idea to use quantum processors for simulating quantum systems dates back to Feynman \cite{Feynman1982} and was later formalized by Lloyd \cite{Lloyd1996}, who together with Abrams, also developed the first algorithms for simulating fermions \cite{Abrams1997}. The idea to use such simulations to prepare ground states in quantum chemistry was proposed by Aspuru-Guzik \emph{et al.} \cite{Aspuru-Guzik2005}.

That original work simulated the quantum chemistry Hamiltonian in a Gaussian orbital basis. While Gaussian orbitals are compact for molecules, they lead to complex Hamiltonians. Initial approaches had gate complexity ${\cal O}(N^{10})$ \cite{Whitfield2010,Wecker2014}, and the current lowest scaling algorithm in that representation has gate complexity of roughly $\widetilde{\cal O}(N^4)$\footnote{Throughout this paper we use the notation $\widetilde{\cal O}(\cdot)$ to indicate an asymptotic upper bound suppressing polylogarithmic factors.} \cite{Berry2019}, where $N$ is number of Gaussian orbitals.

Recently, \cite{BabbushLow} showed that using a plane wave basis restores structure to the Hamiltonian which enables more efficient algorithms. Currently, the two best algorithms simulating the plane wave Hamiltonian are one with ${\cal O}(N)$ spatial complexity and ${\cal O}(N^3)$ gate complexity (with small constant factors) \cite{BabbushSpectra} and one with ${\cal O}(N \log N)$ spatial complexity and $\widetilde{\cal O}(N^2)$ gate complexity (with large constant factors) \cite{Low2018}, where $N$ is the number of plane waves. The scaling of references \cite{BabbushSpectra} and \cite{Low2018} assumes constant resolution and volume proportional to $N$. However, a more appropriate assumption when studying molecules is to take volume proportional to the number of electrons $\eta$, in which case the method of \cite{BabbushSpectra} yields gate complexity ${\cal O}(N^{10/3}/\eta^{1/3} + N^{8/3}/\eta^{2/3})$ and \cite{Low2018} yields gate complexity $\widetilde{\cal O}(N^{8/3}/\eta^{2/3})$. The reason for taking volume proportional to $\eta$ is that for condensed-phase systems (e.g., periodic materials) the electron density is independent of computational cell volume and for single molecules the wavefunction dies off exponentially in space outside of a volume scaling linearly in $\eta$.

While basis set discretization error is suppressed asymptotically as ${\cal O}(1/N)$ regardless of whether $N$ is the number of plane waves \cite{Harl2008,shepherd2012convergence} or Gaussians \cite{helgaker1997basis,Halkier1998}, there is a significant constant factor difference. Plane waves are the standard for treating periodic systems but one needs roughly a hundred times more plane waves than Gaussians \cite{BabbushLow} to reach the accuracy needed to predict chemical reaction rates. Since requiring a hundred times more qubits is impractical in most contexts, this limits the applicability of these recent algorithms \cite{BabbushLow,Kivlichan2017,Low2018,Kivlichan2019,BabbushSpectra} for molecules.

This work solves the plane wave resolution problem by introducing an algorithm with ${\cal O}(\eta \log N)$ spatial complexity and $\widetilde{\cal O}(N^{1/3}\eta^{8/3})$ gate complexity.
With this sublinear scaling in $N$, one can perform simulations with a huge number of plane waves at relatively low cost. Our approach is based on simulating a first-quantized momentum space representation of the potential from the rotating frame of the kinetic operator by using recently introduced interaction picture simulation techniques \cite{Low2018}. While the actual implementations have little in common, our algorithm is conceptually dual to the interaction picture work of \cite{Low2018} which simulates a second-quantized plane wave dual representation of the kinetic operator from the rotating frame of the potential. It is also possible to achieve sublinear scaling in basis size without the interaction picture technique; we briefly discuss how qubitization \cite{Low2016} could be used to obtain $\widetilde{\cal O}(\eta^{4/3}N^{2/3} + \eta^{8/3}N^{1/3})$ scaling.

The reason for our greatly increased efficiency is that the complexity scales like the maximum possible energy representable in the basis.
In second quantization, the basis includes states with up to $\eta=N$ electrons, which results in very high energy, even though these states are not used in the simulation.
In contrast, first quantization sets the number of electrons at $\eta$.
There is still some polynomial scaling with $N$, which is a result of the increased resolution meaning that electrons can be closer together.
If we were to consider constant resolution (as in \cite{BabbushLow,BabbushSpectra,Low2018}), our complexity would actually be polylogarithmic in $N$, representing an exponential speedup in basis size.

\section{Results}

\subsection{Encoding Quantum Simulations of Electronic Structure in Momentum Space First Quantization}

We will represent our system of $\eta$ particles in $N$ orbitals using first quantization. Thus, we require $\eta$ registers (one for each particle) of size $\log N$ (indexing which orbitals are occupied). Since electrons are antisymmetric our registers will encode the wavefunction as
\begin{align}
\label{eq:first_quantized_wavefunction}
\ket{\psi} & = \sum_{p_\ell \in \set} \alpha_{p_1 \cdots p_{\eta}} \ket{p_1 \cdots p_i  \cdots p_j  \cdots p_{\eta}}\nonumber\\
&  =  -  \sum_{p_\ell \in \set}  \alpha_{p_1 \cdots p_{\eta}} \ket{p_1 \cdots p_j  \cdots p_i  \cdots p_{\eta}}
\end{align}
where $\set$ is a set of $N$ spin-orbitals,
so the summation goes over all subsets of the orbitals that contain $\eta$ unique elements. The second line of the above equation is being used to convey that, due to antisymmetry, swapping any two electron registers induces a phase of $-1$. We will specialize to plane wave orbitals in three dimensions and ignore the spin for simplicity, so $\set = [-N^{1/3}/2, N^{1/3}/2]^{3} \subset \mathbb{Z}^3$.
Using plane waves,
\begin{equation}
\label{eq:computational_basis_state}
\braket{r_1 \cdots r_{\eta}}{p_1 \cdots p_{\eta}} \equiv \sqrt{\frac{1}{\Omega^{\eta}}} \prod_{j=1}^\eta e^{-i \, k_{p_{j}} \cdot r_j} 
\end{equation}
where $r_j$ is the position of electron $j$ in real space, $\Omega$ is the computational cell volume, and $k_p = 2 \pi p / \Omega^{1/3}$ is the wavenumber of plane wave $p$.

Unlike in second quantization where antisymmetry is enforced in the operators so that product states of qubits encode Slater determinants \cite{Ortiz2001,Bravyi2002,Seeley2012,Setia2017,Bravyi2017,Jiang2018}, the antisymmetrization indicated in the second line of \eq{first_quantized_wavefunction} must be enforced explicitly in the wavefunction since the computational basis states in \eq{computational_basis_state} are not antisymmetric. However, any initial state can be antisymmetrized with gate complexity ${\cal O}(\eta\log\eta \log N)$ using the techniques recently introduced in \cite{Berry2018}. Evolution under the Hamiltonian will maintain antisymmetry provided that it exists in the initial state (a consequence of fermionic Hamiltonians commuting with the electron permutation operator).

The use of first quantization dates back to the earliest work in quantum simulation \cite{Lloyd1996,Boghosian1998,Zalka1998,Abrams1997,Lidar1999}. Though less common for fermionic systems, several papers have analyzed chemistry simulations using first quantization of real space grids \cite{Kassal2008,Ward2009,Kivlichan2016}. Such grids are incompatible with a Galerkin formulation (the usual discretization strategy used in chemistry involving integrals over the basis) and require methods such as finite-difference discretization, which lack the variational bounds on basis error guaranteed by the Galerkin formulation. Real space grids also have different convergence properties; for example, \cite{Kivlichan2016} finds that in order to maintain constant precision in the representation of certain states, the inverse grid spacing must sometimes scale exponentially in particle number.

Two previous papers \cite{Toloui2013,BabbushSparse2} have presented simulation algorithms within a Gaussian orbital basis at spatial complexity ${\cal O}(\eta \log N)$. These approaches do not use first quantization (they still enforce symmetry in the operators rather than in the wavefunction); instead, \cite{Toloui2013,BabbushSparse2} simulate a block of fixed particle number in the second-quantized Hamiltonian known as the configuration interaction matrix. The more efficient of these two approaches has $\widetilde{\cal O}(\eta^2 N^3)$ gate complexity  \cite{BabbushSparse2}, so our $\widetilde{\cal O}(\eta^{8/3} N^{1/3})$ gate complexity is a substantial improvement.

By integrating the plane wave basis functions with the Laplacian and Coulomb operators in the usual Galerkin formulation \cite{Martin2004} we obtain $H = T + U + V$ such that
\begin{align}
\label{eq:momentum_space}
T & = \sum_{j=1}^{\eta} \sum_{p \in \set} \frac{\|k_p\|^2}{2} \proj{p}_j\\
\label{eq:momentum_space_u}
U & =- \frac{4 \pi}{\Omega} \sum_{\ell=1}^L \sum_{j=1}^{\eta} \sum_{\substack{p,q \in \set \\ p\ne q}} \left(\zeta_\ell \frac{e^{i \, k_{q-p} \cdot R_\ell}}{\|k_{p-q}\|^2}\right)  \ket{p}\!\!\bra{q}_j \\
\label{eq:momentum_space_v}
V & = \frac{2 \pi}{\Omega} \!\sum_{\substack{i,j=1\\ i\ne j}}^{\eta} \sum_{p,q \in \set}  \!\!\!\sum_{\substack{\nu\in \set_0\\ (p+\nu)\in G \\ (q-\nu)\in G}} \!\!\!\!\! \frac{1}{\|k_\nu\|^2} \ket{p+\nu}\!\!\bra{p}_i \cdot \ket{q-\nu}\!\!\bra{q}_j
\end{align}
where $T$ is the kinetic operator, $U$ is the external potential operator, and $V$ is the two-body Coulomb operator.
The set $\set_0$ is $[-N^{1/3}, N^{1/3}]^{3}\backslash \{(0, 0, 0)\} \subset \mathbb{Z}^3$, $R_\ell$ are nuclear coordinates, $\zeta_\ell$ are nuclear charges, $L$ is the number of nuclei, and we use $\ket{q}\!\!\bra{p}_j$ as shorthand notation for
\begin{equation}
I_1 \otimes \cdots \otimes \ket{q}\!\!\bra{p}_j \otimes \cdots \otimes I_{\eta}.
\end{equation}
While this Hamiltonian corresponds to a cubic cell with periodic boundaries, our approach can be easily extended to different lattice geometries (including non-orthogonal unit cells) and systems of reduced periodicity \cite{fusti2002accurate}. Note that we have chosen the frequencies in $G_0$ to span twice the range as those in $G$ in order to cover the maximum momenta that may be exchanged.

\subsection{Simulating Chemistry in the Interaction Picture}

Our scheme for simulation builds on the interaction picture approach introduced in \cite{Low2018}. This approach is useful for performing simulation of a Hamiltonian $H = A + B$ where norms of $A$ and $B$ differ significantly so that $\left \| A \right \| \gg \left\| B \right \|$. The idea is to perform the simulation in the interaction picture in the rotating frame of $A$ so that the large norm of $A$ does not enter the simulation complexity in the usual way.

The principle in \cite{Low2018} is similar to Hamiltonian simulation via a Taylor series \cite{Berry2015}, except that the expression used to approximate the evolution for time $t$ is
\begin{align}
\label{eq:lcu1}
& e^{-i(A+B)t} \approx \sum_{k=0}^{K-1} (-i)^k\int_0^t dt_1 \int_{t_1}^t dt_2 \cdots \int_{t_{k-1}}^t dt_k \, {\cal I}_k\\
&{\cal I}_k = e^{-iA(t-t_k)} B e^{-iA(t_k-t_{k-1})} B \cdots e^{-iA(t_2-t_1)} B e^{-iAt_1}.\nonumber
\end{align}
The operation given by this expression can be implemented by using a linear combination of unitaries (LCU) approach \cite{Childs2012}.
The operator $B$ is expressed as a linear combination of unitaries and the time is discretized, so \eq{lcu1} is a linear combination of unitaries which can be implemented using a control register and oblivious amplitude amplification \cite{Berry2013}.
For a short time, the cutoff $K$ can be chosen logarithmic in the inverse error.
To implement evolution for long times, the time is broken up into a number of time segments of length $\tau$, and this expression is used on each of those segments.

The overall complexity depends on the value of $\lambda$, which is the sum of the weights of the unitaries when expressing $B$ as a sum of unitaries.
To simulate within error $\epsilon$ the number of segments used is ${\cal O}(\lambda t)$, and $K={\cal O}(\log(\lambda t/\epsilon)/\log\log(\lambda t/\epsilon))$.
The complexity in terms of LCU applications of $B$ and evolutions $e^{-iA \tau}$ is therefore
\begin{equation}\label{eq:lamt}
{\cal O}\left( \lambda t\frac{\log(\lambda t/\epsilon)}{\log\log(\lambda t/\epsilon)}\right).
\end{equation}
There is also a multiplicative factor of $\log(t\|A\|/\epsilon\lambda)$ for the gate complexity, which originates from the complexity of preparing the ancilla states used for the time.
This result is given in Lemma 6 of \cite{Low2018}.
To interpret the result as given in \cite{Low2018}, note that the `$\operatorname{HAM-T}$' oracle mentioned in that work includes the evolution under $A$.
That is why the complexity quoted there for the number of applications of $e^{-iA \tau}$ does not include a logarithmic factor.

In quantum chemistry one often decomposes the Hamiltonian into three components $H = T + U + V$, and it is natural to group $U$ and $V$ together, because they usually commute with each other but not with $T$. The work of \cite{Low2018} focused on the simulation of chemistry in second quantization where $\|U + V \| = {\cal O}(N^{7/3}/\Omega^{1/3})$ and $\| T \| = {\cal O}(N^{5/3}/\Omega^{2/3})$, so $\| U + V \| \gg \| T \|$. However, for first-quantized momentum space we will observe the reverse trend that $\| U + V \| \ll \| T \|$ when $N \gg \eta$.

We therefore choose that $A = T$ and $B = U + V$, and need to express the potential as a linear combination of unitaries in momentum space,
\begin{equation}
\label{eq:lcu}
B = U + V = \sum_{s = 1}^{S} w_s H_s \, ,
\qquad
\lambda = \sum_{s=1}^{S}  w_s \, ,
\end{equation}
where $w_s$ are positive scalars and $H_s$ are unitary operators. The convention in this paper is that the $w_s$ are real and non-negative, with any phases included in the $H_s$. A subtlety here is that when expressing $U$ and $V$ as a sum of unitaries we need to account for cases where addition or subtraction with $\nu$ would give values outside $G$. To account for this, we can express $U$ and $V$ as
\begin{align}
\label{eq:lcuUV}
U = & \sum_{\nu \in \set_0} \sum_{\ell=1}^L\frac{ 2 \pi \zeta_\ell  }{\Omega \, \|k_{\nu}\|^2} \sum_{j=1}^{\eta}  \sum_{x\in\{0,1\} } \Bigg( \\
& -e^{-i \, k_{\nu} \cdot R_\ell}\sum_{p \in \set} (-1)^{x[(p-\nu)\notin G]} \ket{p-\nu}\!\!\bra{p}_j\Bigg)\nonumber \\
V = & \sum_{\nu \in \set_0} \frac{\pi}{\Omega \, \|k_\nu\|^2} \sum_{\substack{i,j=1\\i\ne j}}^{\eta} \sum_{x\in\{0,1\} }\Bigg(\\
& \sum_{p,q \in \set} (-1)^{x([(p+\nu)\notin G]\vee[(q-\nu)\notin G])} \ket{p+\nu}\!\!\bra{p}_i \cdot \ket{q-\nu}\!\!\bra{q}_j\Bigg)\nonumber,
\end{align}
and it is apparent that the parts in between the large parentheses above are unitary, so we take them to be the operators $H_s$ in \eq{lcu}. In \eq{lcuUV} we use the convention that Booleans correspond to $0$ for false and $1$ for true. This modification ensures that there is no contribution to the sum from parts where the additions or subtractions would result in values outside $G$.
For example, for $U$, if $(p-\nu)$ is not in $G$, then the value of $(p-\nu)\notin G$ is interpreted as $1$.
This means that we have
\begin{equation}
\label{eq:outside_G}
\sum_{x\in\{0,1\}}(-1)^x=1-1=0.
\end{equation}

Thus, we see that $\lambda$ is asymptotically equal to $\eta^2$ times
\begin{align}
\label{eq:lambda}
\frac{1}{\Omega} \sum_{\nu \in G_0} \frac{\Omega^{2/3}}{\nu^2} &\leq  \frac{1}{\Omega}  \int_0^{2 \pi} \!\! \int_0^{\pi} \int_0^{N^{1/3}} \!\!\!\!\!\! dr \, d\phi \, d\theta \frac{\Omega^{2/3}}{r^2} r^2 \sin \theta \nonumber \\
& = {\cal O}\left(N^{1/3} / \Omega^{1/3}\right) = {\cal O}\left(N^{1/3} / \eta^{1/3}\right)
\end{align}
where in the last line we use $\Omega \propto \eta$, which is typical for molecules \cite{BabbushLow}.
From this, we find that $\lambda = {\cal O}(\eta^{5/3} N^{1/3})$.

The scaling obtained in \eq{lambda} corresponds to the maximum possible value of $U + V$. Changing the orbital basis does not change the maximum possible energy. If one considers the dual basis to the plane waves, then the orbitals are spatially localized with a minimum separation proportional to $(\Omega/N)^{1/3}$ \cite{BabbushLow}. The minimum separation between electrons proportional to $(\Omega/N)^{1/3}$ gives a potential energy scaling as $(N/\Omega)^{1/3}$, as given in \eq{lambda}. In contrast, the kinetic energy $T$ has worse scaling with $N$ because the maximum speed scales as $(N/\Omega)^{1/3}$ (the inverse of the separation), which is squared to give a kinetic energy proportional to $(N/\Omega)^{2/3}$ \cite{BabbushLow}.

\subsection{Performing Simulation in the Kinetic Frame}

To implement our algorithm we need to realize $e^{-i T \tau}$ as well as realize $(U + V) / \lambda$ via a linear combination of unitaries.
Using \eq{momentum_space} we express $e^{-i T \tau}$ as
\begin{equation}
\sum_{p_\ell\in \set} \exp\left[ - \frac{i \tau}2 \sum_{j=1}^{\eta} \|k_{p_j}\|^2\right] \proj{p_1}_1\cdots\proj{p_{\eta}}_{\eta} .
\end{equation}
Therefore, in order to apply this operator, we just need to increment through each of the $\eta$ electron registers to calculate the sum of $\|k_p\|^2$, then apply a phase rotation according to that result.
The complexity of calculating the square $\eta$ times is ${\cal O}(\eta\log^2 N)$ (assuming we are using an elementary multiplication algorithm).
The complexity of the controlled rotations is ${\cal O}(\log(\eta N))$, though there will be an additional logarithmic factor if we consider complexity in terms of T gates for circuit synthesis.

To apply the $U+V$ operator we will need a \textsc{select} operation and a \textsc{prepare} operation.
We use one qubit which selects between performing $U$ and $V$.
For $V$ (the two-electron potential)
the \textsc{select} LCU oracle will be
\begin{align}
&\textsc{select}\ket{0}\ket{i}\ket{j}\ket{\nu} \ket{p_1}_1 \cdots \ket{p_i}_i \cdots \ket{p_j}_j \cdots \ket{p_{\eta}}_{\eta} \mapsto \nonumber\\
& \ket{0}\ket{i}\ket{j}\ket{\nu} \ket{p_1}_1 \! \cdots \ket{p_i + \nu}_i \! \cdots \ket{p_j - \nu}_j \! \cdots \ket{p_{\eta}}_{\eta}.
\end{align}
This operation has complexity ${\cal O}(\eta \log N)$, because we can iterate through each of the $\eta$ electron registers checking if the register number is equal to $i$ or $j$, and if it is then adding $\nu$ (for $i$) or subtracting $\nu$ (for $j$). For $U$ (the nuclear term), we need to apply
\begin{align}
&\textsc{select}\ket{1}\ket{\ell}\ket{j}\ket{\nu} \ket{p_1}_1 \cdots \ket{p_j}_j \cdots \ket{p_{\eta}}_{\eta} \mapsto \nonumber\\
& -e^{-i k_\nu \cdot R_\ell} \ket{1}\ket{\ell}\ket{j}\ket{\nu} \ket{p_1}_1 \cdots \ket{p_j- \nu}_j\cdots \ket{p_{\eta}}_{\eta}.
\end{align}
We again need to iterate through the registers, and subtract $\nu$ if the register number is equal to $j$, which gives complexity ${\cal O}(\eta \log N)$.
The register $\ket{i}$ is replaced with $\ket{\ell}$, and we need to apply a phase factor $e^{-i k_\nu \cdot R_\ell}$.
This phase factor can be obtained by first computing the dot product $k_\nu \cdot R_\ell$, which has complexity ${\cal O}(\log N\log(1/\delta_R))$, where $\delta_R$ is the relative precision with which the positions of the nuclei are specified.
For $L$ nuclei (note that $L \leq \eta$), we will have an additional complexity of ${\cal O}(L\log (1/\delta_R))$ in order to access a classical database for the positions of the nuclei $R_\ell$.
Then, applying the controlled rotation has complexity ${\cal O}(\log N+\log(1/\delta_R))$.

In order to take account of the modification involving $x$ that is referred to in \eq{outside_G}, we would have an additional control qubit for $x$ which would be prepared in an equal superposition. When doing the additions and subtractions, one would check if they give values outside $G$ and perform a $Z$ operation on that ancilla if any of the results were outside $G$.

Let $\delta$ be the allowable error in the \textsc{prepare} and \textsc{select} operations.
The number of times these operations need to be performed is $\widetilde{\cal O}(\lambda t)$, so
to obtain total error no greater than $\epsilon$ we can take $\log(1/\delta)={\cal O}\left(\log\left( \lambda t /\epsilon\right)\right)$.
Since $\lambda$ is polynomial in $\eta$ and $N$, we have $\log(1/\delta)={\cal O}\left(\log\left(\eta N t /\epsilon\right)\right)$.
The error in the implementation of $U/\lambda$ due to the error in the positions of the nuclei is ${\cal O}(\delta_R N^{1/3} Z/\eta)$.
Since the total nuclear charge should be the same as the number of electrons (since the total charge is zero), we can cancel $Z$ and $\eta$.
Then we also obtain $\log(1/\delta_R)={\cal O}\left(\log\left(\eta N t /\epsilon\right)\right)$.

The \textsc{prepare} operation must act as
\begin{align}
&\textsc{prepare}\ket{0}^{\otimes (\log N+2\log\eta+1)} \nonumber\\
& \mapsto \left(\ket{0}\sum_{i,j=1}^{\eta} \sum_{\nu \in \set_0} \sqrt{\frac{2 \pi}{\lambda \Omega \, \|k_\nu\|^2}} \ket{\nu}\ket{i}\ket{j} \right. \nonumber \\
& \quad \left. +\ket{1}  \sum_{j=1}^{\eta} \sum_{\ell=1}^L \sum_{\nu\in \set_0} \sqrt{\frac{4\pi \zeta_\ell}{\lambda \Omega \, \|k_\nu\|^2}}\ket{\nu}\ket{\ell}\ket{j}\right).
\end{align}
This state preparation can be performed by initially rotating the first qubit to give the correct weighting between the $U$ and $V$ terms.
We prepare the register $\ket{j}$ in an equal superposition.
If the first qubit is zero (for the $V$ component) we also prepare the penultimate register in an equal superposition over $\ket{i}$.
We do not need to explicitly eliminate the case $i=j$, because in that case the operation performed is the identity and therefore has no effect on the evolution.
Preparing an equal superposition over $\eta$ values has complexity ${\cal O}(\log \eta)$.

In the case that the first qubit is one (for the $U$ component), we need to prepare the penultimate register in a superposition over $\ket{\ell}$ with weightings $\sqrt{\zeta_\ell}$.
The nuclear charges $\zeta_\ell$ will be given by a classical database with complexity ${\cal O}(L)$. To accomplish this one can use the QROM and subsampling strategies discussed in \cite{BabbushSpectra,Low2018a,Berry2019}.  Again recall that $L \leq \eta$. In fact, usually $L \ll \eta$ and the equality is only saturated for systems consistingly entirely of hydrogen atoms. For a material, in practice there will be a limited number of nuclear charges with nuclei in a regular array, so this complexity will instead be logarithmic in $L$. Similarly, for the selected operation, a regular array of nuclei will mean that the complexity of applying the phase factor $e^{-i k_\nu \cdot R_\ell}$ is logarithmic in $L$.

The key difficulty in implementing \textsc{prepare} is realizing the superposition over $\nu$ with weightings $1/\|k_\nu\|$.
That is, we aim to prepare a state proportional to
\begin{equation}
\sum_{\nu\in \set_0} \frac 1{\|k_\nu\|} \ket{\nu}.
\label{eq:momentum_state}
\end{equation}
We describe this procedure in the next section. If there is a regular array of nuclei then the overall complexity obtained is ${\cal O}(\log (\eta N t/\epsilon)\log N)$, where we use the fact that $L < \eta$. If a full classical database for the nuclei is required, then the complexity will have an additional factor of ${\cal O}(L\log (\eta N t/\epsilon))$.

Between implementing $e^{-iT \tau}$, \textsc{select}, and \textsc{prepare}, the dominant cost is ${\cal O}(\eta \log^2 N)$ for implementing $e^{-iT \tau}$.
There is also a cost of ${\cal O}(\log (\eta N t/\epsilon)\log N)$ for computing $k_\nu \cdot R_\ell$, but in practice it should be smaller.
The factor of $\log(t\|A\|/\epsilon\lambda)$ from \cite{Low2018} will also be smaller.
These are the costs of a single segment, and the number of segments
is given by \eq{lamt} as $\widetilde{\cal O}(\lambda t)$, with $\lambda={\cal O}(\eta^{5/3} N^{1/3})$.
Thus, the total complexity is $\widetilde{\cal O}(\eta^{8/3} N^{1/3} t)$.

\subsection{Preparing the Momentum State}

In this section we develop a surprisingly efficient algorithm for preparing the state in \eq{momentum_state}. Since this step would otherwise be the bottleneck of our algorithm, our implementation is crucial for the overall scaling. The general approach is to use a series of larger and larger nested cubes, each of which is larger than the previous by a factor of $2$.
The index $\cub$ controls which cube we consider.
For each $\cub$ we prepare a set of $\nu$ values in that cube.
We initially prepare a superposition state
\begin{equation}
\frac 1{\sqrt{2^{n+1}-4}}\sum_{\cub=2}^\cubes \sqrt{2^\cub} \ket{\cub}
\end{equation}
which ensures that we obtain the correct weighting for each cube.
This state may be prepared with complexity ${\cal O}(\cubes)$, which is low cost because $\cubes$ is logarithmic in $N$.
The overall preparation will be efficient since the value of $1/\|k_\nu\|$ does not vary by a large amount within each cube, so the amplitude for success is large.
The variation of $1/\|k_\nu\|$ between cubes is accounted for by weighting in the initial superposition over $\cub$.

To simplify the description of the state preparation, we assume that the representation of the integers for $\nu$ uses sign bits.
The sign bits will need to be taken account of in the addition circuits.
It also needs to be taken account of in the preparation, because there are two distinct combinations that correspond to zero.
If each $\nu_x$, $\nu_y$, and $\nu_z$ is represented by $\cubes$ bits, then each will give numbers from $-(2^{\cubes-1}-1)$ to $2^{\cubes-1}-1$.
That is, we have $N^{1/3}=2^{\cubes-1}-1$.
Controlled by $\cub$ we perform Hadamards on $\cub$ of the qubits representing $\nu_x,\nu_y,\nu_z$ to represent the values from $-(2^{\cub-1}-1)$ to $2^{\cub-1}-1$.

As mentioned above, due to the representation of the integers the number zero is represented twice, with a plus sign and a minus sign.
To ensure that all numbers have the same weighting at this stage, we will flag a minus zero as a failure.
The total number of combinations before flagging the failure is $2^{3\cub}$ so the squared amplitude is the inverse of this.
Therefore, the state at this stage is
\begin{equation}
\frac 1{\sqrt{2^{\cubes+1}-4}}\sum_{\cub=2}^\cubes \sum_{\nu_x,\nu_y,\nu_z=-(2^{\cub-1}-1)}^{2^{\cub-1}-1} \!\!\!\!\!\!\!\!\!\!2^{-\cub}\ket{\cub}\ket{\nu_x}\ket{\nu_y}\ket{\nu_z}.
\end{equation}

Next, we test whether \emph{all} of $\nu_x,\nu_y,\nu_z$ are smaller than (in absolute value) $2^{\cub-2}$.
If they are, then the point is inside the box for the next lower value of $\cub$, and we flag failure on an ancilla qubit.
Note that for $\cub=2$ this means that we test whether $\nu=0$, which we need to omit.
This requires testing if all of three bits for $\nu_x,\nu_y,\nu_z$ are zero.
The three bits that are tested depend on $\cub$, so the complexity is ${\cal O}(\cubes)$ (due to the need to check all $3\cubes$ qubits).
The state excluding the failures can then be given as
\begin{equation}
\frac 1{\sqrt{2^{\cubes+1}-4}}\sum_{\cub=2}^\cubes \sum_{\nu\in B_\cub} \frac 1{2^\cub} \ket{\cub}\ket{\nu_x}\ket{\nu_y}\ket{\nu_z},
\end{equation}
where $B_\cub$ (for box $\cub$) is the set of $\nu$ such that the absolute values of $\nu_x,\nu_y,\nu_z$ are less than $2^{\cub-1}$, but it is not the case that they are all less than $2^{\cub-2}$. That is,
\begin{align}
B_\cub = \{ \nu | & (0\le|\nu_x|<2^{\cub-1}) \land (0\le|\nu_y|<2^{\cub-1})\\
& \land (0\le|\nu_z|<2^{\cub-1}) \land ((|\nu_x| \mathbin{\ge} 2^{\cub-2})\nonumber\\
& \lor (|\nu_y| \mathbin{\ge} 2^{\cub-2}) \lor (|\nu_z| \mathbin{\ge} 2^{\cub-2})) \}.\nonumber
\end{align}

Next we prepare an ancilla register in an equal superposition of $\ket{m}$ for $m=0$ to $M-1$, where $M$ is a power of two and is chosen to be large enough to provide a sufficiently accurate approximation of the overall state preparation.
The preparation of the superposition for $m$ can be obtained entirely using Hadamards.
We test the inequality
\begin{equation}
(2^{\cub-2}/\|\nu\|)^2 > m/M \, .
\end{equation}
The left-hand side can be as large as 1 in this region, because we can have just one of $\nu_x,\nu_y,\nu_z$ as large as $2^{\cub-2}$, and the other two equal to zero.
That is, we are at the center of a face of the inner cube.
In order to avoid divisions which are costly to implement, the inequality testing will be performed as
\begin{equation}
(2^{\cub-2})^2 M > m (\nu_x^2+\nu_y^2+\nu_z^2) \, .
\end{equation}

The resulting state will be (omitting the parts where the inequality is not satisfied)
\begin{align}
&\frac 1{\sqrt{M(2^{\cubes+1}-4)}}\sum_{\cub=2}^\cubes \sum_{\nu\in B_\cub}  \sum_{m=0}^{Q-1}  \frac 1{2^\cub} \ket{\cub}\ket{\nu_x}\ket{\nu_y}\ket{\nu_z}\ket{m} \, ,
\end{align}
where $Q = \lceil M (2^{\cub-2}/\|\nu\|)^2 \rceil$ is the number of values of $m$ satisfying the inequality.
The amplitude for each $\nu$ will then be proportional to the square root of the number of values of $m$, and so is
\begin{equation}
\sqrt{\frac{\lceil M (2^{\cub-2}/\|\nu\|)^2 \rceil}{M 2^{2\cub}(2^{n+1}-4)}} \approx \frac 1{4\sqrt{2^{\cubes+1}-4}} \frac 1{\|\nu\|} \, .
\end{equation}
The two sides are approximately equal for large $M$, and in that limit we obtain amplitudes proportional to $1/\|\nu\|$, as required.
We have omitted the parts of the state flagged as failure, and the norm squared of the success state gives the probability for success.
In the limit of large $M$, the norm squared is
\begin{equation}
P_\cubes=\frac 1{2^5(2^{\cubes}-2)} \sum_{\substack{\nu_x,\nu_y,\nu_z=-(2^{\cubes-1}-1)\\ \nu\ne 0}}^{2^{\cubes-1}-1} \frac 1{\nu_x^2+\nu_y^2+\nu_z^2} \, .
\end{equation}

In the limit of large $n$, this expression can be approximated by an integral, which gives
\begin{equation}
P_\cubes\approx\frac 18 \int_0^1 dx \int_0^1 dy \int_0^1 dz \frac 1{x^2+y^2+z^2}.
\end{equation}
The integral is the box integral $B_3(-2)$ in \cite{Bailey2010}.
In \cite{Bailey2010}, $B_3(-2)=3 C_2(-2,1)$, and $C_2(-2,1)$ is given by Eq.~(40) of that work with $a=1$.
That gives the asymptotic value
\begin{align}
P_\cubes &\approx\frac 38 \left[ {\rm Ti}_2(3-\sqrt{8})-G+\frac\pi 2 \log(1+\sqrt{2}) \right] = 0.2398\ldots ,
\end{align}
where Ti$_2$ is the Lewin inverse tangent integral and $G$ is the Catalan constant.
After a single step of amplitude amplification, the probability of failure is
\begin{align}\label{eq:pfail}
\sin^2 \left(3\arccos\left(\sqrt{P_\cubes}\right)\right)
\approx 0.001261\ldots .
\end{align}

\begin{figure}[tbh]
  \centering
  \includegraphics[width=.85\linewidth]{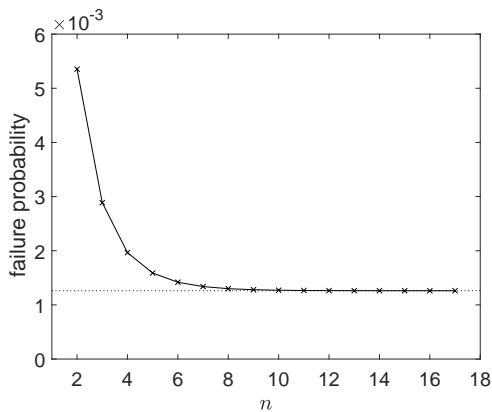}
\caption{The failure probability for the state preparation after a single step of amplitude amplification.
The horizontal dotted line shows the predicted asymptotic value from \eq{pfail}.}
  \label{fig:succ}
\end{figure}

Numerically we find $P_2=11/48$, and it increases with $\cubes$ towards the analytically predicted value.
Using a single step of amplitude amplification brings the probability of success close to $1$ (see \fig{succ}).
Note that the ``failure'' here does not mean that the entire algorithm fails. In the case of ``failure'' of the state preparation one can simply perform the identity instead of the controlled unitaries in the {\sc select} operation. The result is that a small known amount of the identity is added to the Hamiltonian, which can just be subtracted from any eigenvalue estimates. The amplitude amplification triples the complexity for the state preparation.

Next we consider the error in the state preparation due to the finite value of $M$.
The relevant quantity is the sum of the errors in the squared amplitudes, as that gives the error in the weightings of the operations applied to the target state.
That error is upper bounded by
\begin{equation}
\frac 1{M(2^{\cubes+1}-4)}\sum_{\cub=2}^\cubes \sum_{\nu\in B_\cub} \frac 1{2^{2\cub}} < \frac 1{M(2^{\cubes+1}-4)}\sum_{\cub=2}^\cubes {2^{\cub}} = \frac 1{M}.
\end{equation}
This error corresponds to the error in implementation of $(U+V)/\lambda$.
As discussed above, the error in this implementation, $\delta$, can satisfy $\log(1/\delta)={\cal O}(\log(\eta N t/\epsilon))$.
Since $\delta={\cal O}(1/M)$, we can take the number of bits of $M$ as $\log M = {\cal O}(\log(\eta N t/\epsilon))$. The complexities of the steps of this procedure are:
\begin{enumerate}
\item The register $\ket{\mu}$ can be represented in unary, so the state preparation takes a number of gates (rotations and controlled rotations) equal to $\cubes-1={\cal O}(\log N)$, because the dimension is logarithmic in $N$.
\item The superposition over $\nu_x,\nu_y,\nu_z$ can be produced with $3\cubes={\cal O}(\log N)$ controlled Hadamards.
These Hadamards can be controlled by qubits of the unary register used for $\ket{\cub}$.
\item Testing whether the negative zero has been obtained can be performed with a multiply-controlled Toffoli with $\cubes$ controls, which has complexity ${\cal O}(\cubes)={\cal O}(\log N)$.
\item Testing whether the value of $\nu$ is inside the inner box can be performed by using a series of $\cubes$ multiply-controlled Toffolis with $4$ controls (with a unary qubit for $\ket{\cub}$), and one qubit each from the registers for each of the components of $\nu$.
The complexity is therefore ${\cal O}(\cubes)={\cal O}(\log N)$.
\item The preparation of the equal superposition over $m$ has complexity ${\cal O}(\log(1/\delta))$ Hadamards.
\item The inequality test involves multiplications, and therefore has complexity given by the product of the number of digits (better-scaling algorithms would only perform better for an unrealistically large number of digits).
The complexity is therefore ${\cal O}(\log(1/\delta)\log N)$.
\end{enumerate}
The inequality test is the most costly step due to the multiplications, and gives the overall cost of the state preparation algorithm.
Nevertheless, it still has logarithmic cost, so the complexity of the \textsc{prepare} operation will be negligible compared to the costs of the other steps.
This concludes our procedure for realizing the state in \eq{momentum_state}, which completes our presentation of the overall simulation procedure.

\section{Conclusion}

The low scaling dependence of our methods on $N$ allows us to easily overcome the constant factor difference in resolution between plane waves and Gaussians. In fact, using these algorithms we expect that one can achieve precisions limited only by relativistic effects and the Born-Oppenheimer approximation. However, the latter limitation can also be alleviated by our approach since one can use enough plane waves to reasonably span the energy scales required for momentum transfer between nuclei and electrons, and thus support simulations with explicit quantum treatment of the nuclei. We also expect that our approach could be viable for the first generation of fault-tolerant quantum computers.

Let us consider the calculation of the FeMoco cofactor of the Nitrogenase enzyme discussed in \cite{Reiher2017,Motta2018,Zhendong2018} which involved 54 electrons and 108 Gaussian spin-orbitals. FeMoco is the active site of biological Nitrogen fixation and its electronic structure has remained elusive to classical methods. The work of \cite{Reiher2017} found that roughly $10^{15}$ T gates would be required, which translates to needing roughly $10^8$ physical qubits if implemented in the surface code with gates at $10^{-3}$ error rate. The large qubit count here arises from needing to parallelize magic state distillation (the system register would need only about $10^5$ physical qubits). In comparison, the ${\cal O}(N^3)$ scaling algorithm of \cite{BabbushSpectra} has been shown to require less than $10^{9}$ T gates to solve a molecule with 100 plane wave spin-orbitals (which is not enough resolution for FeMoco).

Supposing we use $10^6$ plane wave spin-orbitals for these 54 electrons our algorithm would require roughly $10^3$ logical qubits (which can be encoded in roughly $10^6$ physical qubits under the architecture assumptions discussed in \cite{BabbushSpectra}, which are more conservative than those in \cite{Reiher2017}). Under these assumptions the value of $\eta^{8/3} N^{1/3}$ is only about $4 \times 10^6$, though there will be significant logarithmic and constant factors in the gate complexity. In comparison $N^{8/3}/\eta^{2/3}$ for the approach of \cite{Low2018} would be about $7\times 10^{14}$.
While further work would be needed to determine the precise gate counts, it seems reasonable that gate counts would be low enough to perform magic state distillation in series with a single T factory. This back-of-the-envelope estimate suggests that our approach could surpass the accuracy of the FeMoco simulation discussed in \cite{Reiher2017} while using fewer physical qubits. Using such a large basis may also alleviate the need for active space perturbation techniques such as those discussed in \cite{Takeshita2019}.

Although we have used the interaction picture to achieve our $\widetilde{\cal O}(\eta^{8/3} N^{1/3} t)$ complexity, it is also possible to achieve sublinear complexity in $N$ without the interaction picture. This is because the value of $\lambda$ associated with the kinetic operator $T$ is ${\cal O}(\eta^{1/3} N^{2/3})$ \cite{BabbushLow}.
Thus, one could use qubitization and signal processing \cite{Low2016,Low2017} where $T$ is simulated using LCU methods. Then, the overall complexity of our approach would be $\widetilde{\cal O}(\eta^{8/3} N^{1/3} t+\eta^{4/3} N^{2/3}t)$,
and the constant factors in the scaling might be smaller.

\subsection*{Acknowledgements}

The authors thank Ian Kivlichan and Craig Gidney for helpful discussions, as well as Artur Scherer, Yuan Su and Zhang Jiang for feedback on an early draft. D.W.B. is funded by Australian Research Council Discovery projects (Grant Nos.\ DP160102426 and DP190102633).

\subsection*{Author Contributions}

R.B. and D.W.B. jointly developed the algorithms, performed the complexity analysis, and co-wrote the manuscript. J.M. and H.N. assisted with writing and discussions of the first-quantized representation.\\

\bibliography{Mendeley2}

\begin{thebibliography}{44}%
\makeatletter
\providecommand \@ifxundefined [1]{%
 \@ifx{#1\undefined}
}%
\providecommand \@ifnum [1]{%
 \ifnum #1\expandafter \@firstoftwo
 \else \expandafter \@secondoftwo
 \fi
}%
\providecommand \@ifx [1]{%
 \ifx #1\expandafter \@firstoftwo
 \else \expandafter \@secondoftwo
 \fi
}%
\providecommand \natexlab [1]{#1}%
\providecommand \enquote  [1]{``#1''}%
\providecommand \bibnamefont  [1]{#1}%
\providecommand \bibfnamefont [1]{#1}%
\providecommand \citenamefont [1]{#1}%
\providecommand \href@noop [0]{\@secondoftwo}%
\providecommand \href [0]{\begingroup \@sanitize@url \@href}%
\providecommand \@href[1]{\@@startlink{#1}\@@href}%
\providecommand \@@href[1]{\endgroup#1\@@endlink}%
\providecommand \@sanitize@url [0]{\catcode `\\12\catcode `\$12\catcode
  `\&12\catcode `\#12\catcode `\^12\catcode `\_12\catcode `\%12\relax}%
\providecommand \@@startlink[1]{}%
\providecommand \@@endlink[0]{}%
\providecommand \url  [0]{\begingroup\@sanitize@url \@url }%
\providecommand \@url [1]{\endgroup\@href {#1}{\urlprefix }}%
\providecommand \urlprefix  [0]{URL }%
\providecommand \Eprint [0]{\href }%
\providecommand \doibase [0]{http://dx.doi.org/}%
\providecommand \selectlanguage [0]{\@gobble}%
\providecommand \bibinfo  [0]{\@secondoftwo}%
\providecommand \bibfield  [0]{\@secondoftwo}%
\providecommand \translation [1]{[#1]}%
\providecommand \BibitemOpen [0]{}%
\providecommand \bibitemStop [0]{}%
\providecommand \bibitemNoStop [0]{.\EOS\space}%
\providecommand \EOS [0]{\spacefactor3000\relax}%
\providecommand \BibitemShut  [1]{\csname bibitem#1\endcsname}%
\let\auto@bib@innerbib\@empty
\bibitem [{\citenamefont {Feynman}(1982)}]{Feynman1982}%
  \BibitemOpen
  \bibfield  {author} {\bibinfo {author} {\bibfnamefont {Richard~P}\
  \bibnamefont {Feynman}},\ }\bibfield  {title} {\enquote {\bibinfo {title}
  {{Simulating physics with computers}},}\ }\href {\doibase 10.1007/BF02650179}
  {\bibfield  {journal} {\bibinfo  {journal} {International Journal of
  Theoretical Physics}\ }\textbf {\bibinfo {volume} {21}},\ \bibinfo {pages}
  {467--488} (\bibinfo {year} {1982})}\BibitemShut {NoStop}%
\bibitem [{\citenamefont {Lloyd}(1996)}]{Lloyd1996}%
  \BibitemOpen
  \bibfield  {author} {\bibinfo {author} {\bibfnamefont {Seth}\ \bibnamefont
  {Lloyd}},\ }\bibfield  {title} {\enquote {\bibinfo {title} {Universal quantum
  simulators},}\ }\href {\doibase 10.1126/science.273.5278.1073} {\bibfield
  {journal} {\bibinfo  {journal} {Science}\ }\textbf {\bibinfo {volume}
  {273}},\ \bibinfo {pages} {1073--1078} (\bibinfo {year} {1996})}\BibitemShut
  {NoStop}%
\bibitem [{\citenamefont {Abrams}\ and\ \citenamefont
  {Lloyd}(1997)}]{Abrams1997}%
  \BibitemOpen
  \bibfield  {author} {\bibinfo {author} {\bibfnamefont {Daniel~S}\
  \bibnamefont {Abrams}}\ and\ \bibinfo {author} {\bibfnamefont {Seth}\
  \bibnamefont {Lloyd}},\ }\bibfield  {title} {\enquote {\bibinfo {title}
  {Simulation of many-body {F}ermi systems on a universal quantum computer},}\
  }\href {https://doi.org/10.1103/PhysRevLett.79.2586} {\bibfield  {journal}
  {\bibinfo  {journal} {Physical Review Letters}\ }\textbf {\bibinfo {volume}
  {79}},\ \bibinfo {pages} {2586--2589} (\bibinfo {year} {1997})}\BibitemShut
  {NoStop}%
\bibitem [{\citenamefont {Aspuru-Guzik}\ \emph {et~al.}(2005)\citenamefont
  {Aspuru-Guzik}, \citenamefont {Dutoi}, \citenamefont {Love},\ and\
  \citenamefont {Head-Gordon}}]{Aspuru-Guzik2005}%
  \BibitemOpen
  \bibfield  {author} {\bibinfo {author} {\bibfnamefont {Al{\'a}n}\
  \bibnamefont {Aspuru-Guzik}}, \bibinfo {author} {\bibfnamefont {Anthony~D}\
  \bibnamefont {Dutoi}}, \bibinfo {author} {\bibfnamefont {Peter~J}\
  \bibnamefont {Love}}, \ and\ \bibinfo {author} {\bibfnamefont {Martin}\
  \bibnamefont {Head-Gordon}},\ }\bibfield  {title} {\enquote {\bibinfo {title}
  {Simulated quantum computation of molecular energies},}\ }\href {\doibase
  10.1126/science.1113479} {\bibfield  {journal} {\bibinfo  {journal}
  {Science}\ }\textbf {\bibinfo {volume} {309}},\ \bibinfo {pages} {1704}
  (\bibinfo {year} {2005})}\BibitemShut {NoStop}%
\bibitem [{\citenamefont {Whitfield}\ \emph {et~al.}(2011)\citenamefont
  {Whitfield}, \citenamefont {Biamonte},\ and\ \citenamefont
  {Aspuru-Guzik}}]{Whitfield2010}%
  \BibitemOpen
  \bibfield  {author} {\bibinfo {author} {\bibfnamefont {James~D}\ \bibnamefont
  {Whitfield}}, \bibinfo {author} {\bibfnamefont {Jacob}\ \bibnamefont
  {Biamonte}}, \ and\ \bibinfo {author} {\bibfnamefont {Al{\'a}n}\ \bibnamefont
  {Aspuru-Guzik}},\ }\bibfield  {title} {\enquote {\bibinfo {title}
  {{Simulation of electronic structure Hamiltonians using quantum
  computers}},}\ }\href {\doibase 10.1080/00268976.2011.552441} {\bibfield
  {journal} {\bibinfo  {journal} {Molecular Physics}\ }\textbf {\bibinfo
  {volume} {109}},\ \bibinfo {pages} {735--750} (\bibinfo {year}
  {2011})}\BibitemShut {NoStop}%
\bibitem [{\citenamefont {Wecker}\ \emph {et~al.}(2014)\citenamefont {Wecker},
  \citenamefont {Bauer}, \citenamefont {Clark}, \citenamefont {Hastings},\ and\
  \citenamefont {Troyer}}]{Wecker2014}%
  \BibitemOpen
  \bibfield  {author} {\bibinfo {author} {\bibfnamefont {David}\ \bibnamefont
  {Wecker}}, \bibinfo {author} {\bibfnamefont {Bela}\ \bibnamefont {Bauer}},
  \bibinfo {author} {\bibfnamefont {Bryan~K}\ \bibnamefont {Clark}}, \bibinfo
  {author} {\bibfnamefont {Matthew~B}\ \bibnamefont {Hastings}}, \ and\
  \bibinfo {author} {\bibfnamefont {Matthias}\ \bibnamefont {Troyer}},\
  }\bibfield  {title} {\enquote {\bibinfo {title} {{Gate-count estimates for
  performing quantum chemistry on small quantum computers}},}\ }\href {\doibase
  10.1103/PhysRevA.90.022305} {\bibfield  {journal} {\bibinfo  {journal}
  {Physical Review A}\ }\textbf {\bibinfo {volume} {90}},\ \bibinfo {pages}
  {022305} (\bibinfo {year} {2014})}\BibitemShut {NoStop}%
\bibitem [{\citenamefont {Berry}\ \emph {et~al.}(2019)\citenamefont {Berry},
  \citenamefont {Gidney}, \citenamefont {Motta}, \citenamefont {McClean},\ and\
  \citenamefont {Babbush}}]{Berry2019}%
  \BibitemOpen
  \bibfield  {author} {\bibinfo {author} {\bibfnamefont {Dominic~W}\
  \bibnamefont {Berry}}, \bibinfo {author} {\bibfnamefont {Craig}\ \bibnamefont
  {Gidney}}, \bibinfo {author} {\bibfnamefont {Mario}\ \bibnamefont {Motta}},
  \bibinfo {author} {\bibfnamefont {Jarrod}\ \bibnamefont {McClean}}, \ and\
  \bibinfo {author} {\bibfnamefont {Ryan}\ \bibnamefont {Babbush}},\ }\bibfield
   {title} {\enquote {\bibinfo {title} {Qubitization of arbitrary basis quantum
  chemistry by low rank factorization},}\ }\href
  {https://arxiv.org/abs/1902.02134} {\bibfield  {journal} {\bibinfo  {journal}
  {arXiv:1902.02134}\ } (\bibinfo {year} {2019})}\BibitemShut {NoStop}%
\bibitem [{\citenamefont {Babbush}\ \emph
  {et~al.}(2018{\natexlab{a}})\citenamefont {Babbush}, \citenamefont {Wiebe},
  \citenamefont {McClean}, \citenamefont {McClain}, \citenamefont {Neven},\
  and\ \citenamefont {Chan}}]{BabbushLow}%
  \BibitemOpen
  \bibfield  {author} {\bibinfo {author} {\bibfnamefont {Ryan}\ \bibnamefont
  {Babbush}}, \bibinfo {author} {\bibfnamefont {Nathan}\ \bibnamefont {Wiebe}},
  \bibinfo {author} {\bibfnamefont {Jarrod}\ \bibnamefont {McClean}}, \bibinfo
  {author} {\bibfnamefont {James}\ \bibnamefont {McClain}}, \bibinfo {author}
  {\bibfnamefont {Hartmut}\ \bibnamefont {Neven}}, \ and\ \bibinfo {author}
  {\bibfnamefont {Garnet Kin-Lic}\ \bibnamefont {Chan}},\ }\bibfield  {title}
  {\enquote {\bibinfo {title} {Low-depth quantum simulation of materials},}\
  }\href {https://journals.aps.org/prx/abstract/10.1103/PhysRevX.8.011044}
  {\bibfield  {journal} {\bibinfo  {journal} {Physical Review X}\ }\textbf
  {\bibinfo {volume} {8}},\ \bibinfo {pages} {011044} (\bibinfo {year}
  {2018}{\natexlab{a}})}\BibitemShut {NoStop}%
\bibitem [{\citenamefont {Babbush}\ \emph
  {et~al.}(2018{\natexlab{b}})\citenamefont {Babbush}, \citenamefont {Gidney},
  \citenamefont {Berry}, \citenamefont {Wiebe}, \citenamefont {McClean},
  \citenamefont {Paler}, \citenamefont {Fowler},\ and\ \citenamefont
  {Neven}}]{BabbushSpectra}%
  \BibitemOpen
  \bibfield  {author} {\bibinfo {author} {\bibfnamefont {Ryan}\ \bibnamefont
  {Babbush}}, \bibinfo {author} {\bibfnamefont {Craig}\ \bibnamefont {Gidney}},
  \bibinfo {author} {\bibfnamefont {Dominic~W}\ \bibnamefont {Berry}}, \bibinfo
  {author} {\bibfnamefont {Nathan}\ \bibnamefont {Wiebe}}, \bibinfo {author}
  {\bibfnamefont {Jarrod}\ \bibnamefont {McClean}}, \bibinfo {author}
  {\bibfnamefont {Alexandru}\ \bibnamefont {Paler}}, \bibinfo {author}
  {\bibfnamefont {Austin}\ \bibnamefont {Fowler}}, \ and\ \bibinfo {author}
  {\bibfnamefont {Hartmut}\ \bibnamefont {Neven}},\ }\bibfield  {title}
  {\enquote {\bibinfo {title} {Encoding electronic spectra in quantum circuits
  with linear {T} complexity},}\ }\href
  {https://journals.aps.org/prx/abstract/10.1103/PhysRevX.8.041015} {\bibfield
  {journal} {\bibinfo  {journal} {Physical Review X}\ }\textbf {\bibinfo
  {volume} {8}},\ \bibinfo {pages} {041015} (\bibinfo {year}
  {2018}{\natexlab{b}})}\BibitemShut {NoStop}%
\bibitem [{\citenamefont {Low}\ and\ \citenamefont {Wiebe}(2018)}]{Low2018}%
  \BibitemOpen
  \bibfield  {author} {\bibinfo {author} {\bibfnamefont {Guang~Hao}\
  \bibnamefont {Low}}\ and\ \bibinfo {author} {\bibfnamefont {Nathan}\
  \bibnamefont {Wiebe}},\ }\bibfield  {title} {\enquote {\bibinfo {title}
  {Hamiltonian simulation in the interaction picture},}\ }\href
  {http://arxiv.org/abs/1805.00675} {\bibfield  {journal} {\bibinfo  {journal}
  {arXiv:1805.00675}\ } (\bibinfo {year} {2018})}\BibitemShut {NoStop}%
\bibitem [{\citenamefont {Harl}\ and\ \citenamefont {Kresse}(2008)}]{Harl2008}%
  \BibitemOpen
  \bibfield  {author} {\bibinfo {author} {\bibfnamefont {Judith}\ \bibnamefont
  {Harl}}\ and\ \bibinfo {author} {\bibfnamefont {Georg}\ \bibnamefont
  {Kresse}},\ }\bibfield  {title} {\enquote {\bibinfo {title} {{Cohesive energy
  curves for noble gas solids calculated by adiabatic connection
  fluctuation-dissipation theory}},}\ }\href {\doibase
  10.1103/PhysRevB.77.045136} {\bibfield  {journal} {\bibinfo  {journal}
  {Physical Review B}\ }\textbf {\bibinfo {volume} {77}},\ \bibinfo {pages}
  {45136} (\bibinfo {year} {2008})}\BibitemShut {NoStop}%
\bibitem [{\citenamefont {Shepherd}\ \emph {et~al.}(2012)\citenamefont
  {Shepherd}, \citenamefont {Gr{\"{u}}neis}, \citenamefont {Booth},
  \citenamefont {Kresse},\ and\ \citenamefont
  {Alavi}}]{shepherd2012convergence}%
  \BibitemOpen
  \bibfield  {author} {\bibinfo {author} {\bibfnamefont {James~J}\ \bibnamefont
  {Shepherd}}, \bibinfo {author} {\bibfnamefont {Andreas}\ \bibnamefont
  {Gr{\"{u}}neis}}, \bibinfo {author} {\bibfnamefont {George~H}\ \bibnamefont
  {Booth}}, \bibinfo {author} {\bibfnamefont {Georg}\ \bibnamefont {Kresse}}, \
  and\ \bibinfo {author} {\bibfnamefont {Ali}\ \bibnamefont {Alavi}},\
  }\bibfield  {title} {\enquote {\bibinfo {title} {{Convergence of many-body
  wave-function expansions using a plane-wave basis: From homogeneous electron
  gas to solid state systems}},}\ }\href {\doibase 10.1103/PhysRevB.86.035111}
  {\bibfield  {journal} {\bibinfo  {journal} {Physical Review B}\ }\textbf
  {\bibinfo {volume} {86}},\ \bibinfo {pages} {35111} (\bibinfo {year}
  {2012})}\BibitemShut {NoStop}%
\bibitem [{\citenamefont {Helgaker}\ \emph {et~al.}(1998)\citenamefont
  {Helgaker}, \citenamefont {Klopper}, \citenamefont {Koch},\ and\
  \citenamefont {Noga}}]{helgaker1997basis}%
  \BibitemOpen
  \bibfield  {author} {\bibinfo {author} {\bibfnamefont {Trygve}\ \bibnamefont
  {Helgaker}}, \bibinfo {author} {\bibfnamefont {Wim}\ \bibnamefont {Klopper}},
  \bibinfo {author} {\bibfnamefont {Henrik}\ \bibnamefont {Koch}}, \ and\
  \bibinfo {author} {\bibfnamefont {Jozef}\ \bibnamefont {Noga}},\ }\bibfield
  {title} {\enquote {\bibinfo {title} {{Basis-set convergence of correlated
  calculations on water}},}\ }\href {\doibase 10.1063/1.473863} {\bibfield
  {journal} {\bibinfo  {journal} {Journal of Chemical Physics}\ }\textbf
  {\bibinfo {volume} {106}},\ \bibinfo {pages} {9639--9646} (\bibinfo {year}
  {1998})}\BibitemShut {NoStop}%
\bibitem [{\citenamefont {Halkier}\ \emph {et~al.}(1998)\citenamefont
  {Halkier}, \citenamefont {Helgaker}, \citenamefont {J{\o}rgensen},
  \citenamefont {Klopper}, \citenamefont {Koch}, \citenamefont {Olsen},\ and\
  \citenamefont {Wilson}}]{Halkier1998}%
  \BibitemOpen
  \bibfield  {author} {\bibinfo {author} {\bibfnamefont {Asger}\ \bibnamefont
  {Halkier}}, \bibinfo {author} {\bibfnamefont {Trygve}\ \bibnamefont
  {Helgaker}}, \bibinfo {author} {\bibfnamefont {Poul}\ \bibnamefont
  {J{\o}rgensen}}, \bibinfo {author} {\bibfnamefont {Wim}\ \bibnamefont
  {Klopper}}, \bibinfo {author} {\bibfnamefont {Henrik}\ \bibnamefont {Koch}},
  \bibinfo {author} {\bibfnamefont {Jeppe}\ \bibnamefont {Olsen}}, \ and\
  \bibinfo {author} {\bibfnamefont {Angela~K}\ \bibnamefont {Wilson}},\
  }\bibfield  {title} {\enquote {\bibinfo {title} {Basis-set convergence in
  correlated calculations on {Ne, N$_2$, and H$_2$O}},}\ }\href {\doibase
  10.1016/S0009-2614(98)00111-0} {\bibfield  {journal} {\bibinfo  {journal}
  {Chemical Physics Letters}\ }\textbf {\bibinfo {volume} {286}},\ \bibinfo
  {pages} {243--252} (\bibinfo {year} {1998})}\BibitemShut {NoStop}%
\bibitem [{\citenamefont {Kivlichan}\ \emph {et~al.}(2018)\citenamefont
  {Kivlichan}, \citenamefont {McClean}, \citenamefont {Wiebe}, \citenamefont
  {Gidney}, \citenamefont {Aspuru-Guzik}, \citenamefont {Chan},\ and\
  \citenamefont {Babbush}}]{Kivlichan2017}%
  \BibitemOpen
  \bibfield  {author} {\bibinfo {author} {\bibfnamefont {Ian~D}\ \bibnamefont
  {Kivlichan}}, \bibinfo {author} {\bibfnamefont {Jarrod}\ \bibnamefont
  {McClean}}, \bibinfo {author} {\bibfnamefont {Nathan}\ \bibnamefont {Wiebe}},
  \bibinfo {author} {\bibfnamefont {Craig}\ \bibnamefont {Gidney}}, \bibinfo
  {author} {\bibfnamefont {Al{\'a}n}\ \bibnamefont {Aspuru-Guzik}}, \bibinfo
  {author} {\bibfnamefont {Garnet Kin-Lic}\ \bibnamefont {Chan}}, \ and\
  \bibinfo {author} {\bibfnamefont {Ryan}\ \bibnamefont {Babbush}},\ }\bibfield
   {title} {\enquote {\bibinfo {title} {Quantum simulation of electronic
  structure with linear depth and connectivity},}\ }\href {\doibase
  10.1103/PhysRevLett.120.110501} {\bibfield  {journal} {\bibinfo  {journal}
  {Physical Review Letters}\ }\textbf {\bibinfo {volume} {120}},\ \bibinfo
  {pages} {110501} (\bibinfo {year} {2018})}\BibitemShut {NoStop}%
\bibitem [{\citenamefont {Kivlichan}\ \emph {et~al.}(2019)\citenamefont
  {Kivlichan}, \citenamefont {Gidney}, \citenamefont {Berry}, \citenamefont
  {Wiebe}, \citenamefont {McClean}, \citenamefont {Sun}, \citenamefont {Jiang},
  \citenamefont {Rubin}, \citenamefont {Fowler}, \citenamefont {Aspuru-Guzik},
  \citenamefont {Neven},\ and\ \citenamefont {Babbush}}]{Kivlichan2019}%
  \BibitemOpen
  \bibfield  {author} {\bibinfo {author} {\bibfnamefont {Ian~D.}\ \bibnamefont
  {Kivlichan}}, \bibinfo {author} {\bibfnamefont {Craig}\ \bibnamefont
  {Gidney}}, \bibinfo {author} {\bibfnamefont {Dominic~W.}\ \bibnamefont
  {Berry}}, \bibinfo {author} {\bibfnamefont {Nathan}\ \bibnamefont {Wiebe}},
  \bibinfo {author} {\bibfnamefont {Jarrod}\ \bibnamefont {McClean}}, \bibinfo
  {author} {\bibfnamefont {Wei}\ \bibnamefont {Sun}}, \bibinfo {author}
  {\bibfnamefont {Zhang}\ \bibnamefont {Jiang}}, \bibinfo {author}
  {\bibfnamefont {Nicholas}\ \bibnamefont {Rubin}}, \bibinfo {author}
  {\bibfnamefont {Austin}\ \bibnamefont {Fowler}}, \bibinfo {author}
  {\bibfnamefont {Al{\'a}n}\ \bibnamefont {Aspuru-Guzik}}, \bibinfo {author}
  {\bibfnamefont {Hartmut}\ \bibnamefont {Neven}}, \ and\ \bibinfo {author}
  {\bibfnamefont {Ryan}\ \bibnamefont {Babbush}},\ }\bibfield  {title}
  {\enquote {\bibinfo {title} {Improved fault-tolerant quantum simulation of
  condensed-phase correlated electrons via {T}rotterization},}\ }\href
  {https://arxiv.org/abs/1902.10673} {\bibfield  {journal} {\bibinfo  {journal}
  {arXiv:1902.10673}\ } (\bibinfo {year} {2019})}\BibitemShut {NoStop}%
\bibitem [{\citenamefont {Low}\ and\ \citenamefont {Chuang}(2019)}]{Low2016}%
  \BibitemOpen
  \bibfield  {author} {\bibinfo {author} {\bibfnamefont {Guang~Hao}\
  \bibnamefont {Low}}\ and\ \bibinfo {author} {\bibfnamefont {Isaac~L}\
  \bibnamefont {Chuang}},\ }\bibfield  {title} {\enquote {\bibinfo {title}
  {Hamiltonian simulation by qubitization},}\ }\href {\doibase
  10.22331/q-2019-07-12-163} {\bibfield  {journal} {\bibinfo  {journal}
  {{Quantum}}\ }\textbf {\bibinfo {volume} {3}},\ \bibinfo {pages} {163}
  (\bibinfo {year} {2019})}\BibitemShut {NoStop}%
\bibitem [{\citenamefont {Ortiz}\ \emph {et~al.}(2001)\citenamefont {Ortiz},
  \citenamefont {Gubernatis}, \citenamefont {Knill},\ and\ \citenamefont
  {Laflamme}}]{Ortiz2001}%
  \BibitemOpen
  \bibfield  {author} {\bibinfo {author} {\bibfnamefont {Gerardo}\ \bibnamefont
  {Ortiz}}, \bibinfo {author} {\bibfnamefont {James~E}\ \bibnamefont
  {Gubernatis}}, \bibinfo {author} {\bibfnamefont {Emanuel}\ \bibnamefont
  {Knill}}, \ and\ \bibinfo {author} {\bibfnamefont {Raymond}\ \bibnamefont
  {Laflamme}},\ }\bibfield  {title} {\enquote {\bibinfo {title} {Quantum
  algorithms for fermionic simulations},}\ }\href {\doibase
  10.1103/PhysRevA.64.022319} {\bibfield  {journal} {\bibinfo  {journal}
  {Physical Review A}\ }\textbf {\bibinfo {volume} {64}},\ \bibinfo {pages}
  {22319} (\bibinfo {year} {2001})}\BibitemShut {NoStop}%
\bibitem [{\citenamefont {Bravyi}\ and\ \citenamefont
  {Kitaev}(2002)}]{Bravyi2002}%
  \BibitemOpen
  \bibfield  {author} {\bibinfo {author} {\bibfnamefont {Sergey}\ \bibnamefont
  {Bravyi}}\ and\ \bibinfo {author} {\bibfnamefont {Alexei}\ \bibnamefont
  {Kitaev}},\ }\bibfield  {title} {\enquote {\bibinfo {title} {{Fermionic
  quantum computation}},}\ }\href {\doibase 10.1006/aphy.2002.6254} {\bibfield
  {journal} {\bibinfo  {journal} {Annals of Physics}\ }\textbf {\bibinfo
  {volume} {298}},\ \bibinfo {pages} {210--226} (\bibinfo {year}
  {2002})}\BibitemShut {NoStop}%
\bibitem [{\citenamefont {Seeley}\ \emph {et~al.}(2012)\citenamefont {Seeley},
  \citenamefont {Richard},\ and\ \citenamefont {Love}}]{Seeley2012}%
  \BibitemOpen
  \bibfield  {author} {\bibinfo {author} {\bibfnamefont {Jacob~T}\ \bibnamefont
  {Seeley}}, \bibinfo {author} {\bibfnamefont {Martin~J}\ \bibnamefont
  {Richard}}, \ and\ \bibinfo {author} {\bibfnamefont {Peter~J}\ \bibnamefont
  {Love}},\ }\bibfield  {title} {\enquote {\bibinfo {title} {{The Bravyi-Kitaev
  transformation for quantum computation of electronic structure}},}\ }\href
  {\doibase 10.1063/1.4768229} {\bibfield  {journal} {\bibinfo  {journal}
  {Journal of Chemical Physics}\ }\textbf {\bibinfo {volume} {137}},\ \bibinfo
  {pages} {224109} (\bibinfo {year} {2012})}\BibitemShut {NoStop}%
\bibitem [{\citenamefont {Setia}\ and\ \citenamefont
  {Whitfield}(2018)}]{Setia2017}%
  \BibitemOpen
  \bibfield  {author} {\bibinfo {author} {\bibfnamefont {Kanav}\ \bibnamefont
  {Setia}}\ and\ \bibinfo {author} {\bibfnamefont {James~D}\ \bibnamefont
  {Whitfield}},\ }\bibfield  {title} {\enquote {\bibinfo {title}
  {{Bravyi-Kitaev Superfast simulation of fermions on a quantum computer}},}\
  }\href {\doibase 10.1063/1.5019371} {\bibfield  {journal} {\bibinfo
  {journal} {The Journal of Chemical Physics}\ }\textbf {\bibinfo {volume}
  {148}},\ \bibinfo {pages} {164104} (\bibinfo {year} {2018})}\BibitemShut
  {NoStop}%
\bibitem [{\citenamefont {Bravyi}\ \emph {et~al.}(2017)\citenamefont {Bravyi},
  \citenamefont {Gambetta}, \citenamefont {Mezzacapo},\ and\ \citenamefont
  {Temme}}]{Bravyi2017}%
  \BibitemOpen
  \bibfield  {author} {\bibinfo {author} {\bibfnamefont {Sergey}\ \bibnamefont
  {Bravyi}}, \bibinfo {author} {\bibfnamefont {Jay~M}\ \bibnamefont
  {Gambetta}}, \bibinfo {author} {\bibfnamefont {Antonio}\ \bibnamefont
  {Mezzacapo}}, \ and\ \bibinfo {author} {\bibfnamefont {Kristan}\ \bibnamefont
  {Temme}},\ }\bibfield  {title} {\enquote {\bibinfo {title} {{Tapering off
  qubits to simulate fermionic Hamiltonians}},}\ }\href
  {http://arxiv.org/abs/1701.08213} {\bibfield  {journal} {\bibinfo  {journal}
  {arXiv:1701.08213}\ } (\bibinfo {year} {2017})}\BibitemShut {NoStop}%
\bibitem [{\citenamefont {Jiang}\ \emph {et~al.}(2018)\citenamefont {Jiang},
  \citenamefont {McClean}, \citenamefont {Babbush},\ and\ \citenamefont
  {Neven}}]{Jiang2018}%
  \BibitemOpen
  \bibfield  {author} {\bibinfo {author} {\bibfnamefont {Zhang}\ \bibnamefont
  {Jiang}}, \bibinfo {author} {\bibfnamefont {Jarrod}\ \bibnamefont {McClean}},
  \bibinfo {author} {\bibfnamefont {Ryan}\ \bibnamefont {Babbush}}, \ and\
  \bibinfo {author} {\bibfnamefont {Hartmut}\ \bibnamefont {Neven}},\
  }\bibfield  {title} {\enquote {\bibinfo {title} {Majorana loop stabilizer
  codes for error correction of fermionic quantum simulations},}\ }\href
  {http://arxiv.org/abs/1812.08190} {\bibfield  {journal} {\bibinfo  {journal}
  {arXiv:1812.08190}\ } (\bibinfo {year} {2018})}\BibitemShut {NoStop}%
\bibitem [{\citenamefont {Berry}\ \emph {et~al.}(2018)\citenamefont {Berry},
  \citenamefont {Kieferov{\'{a}}}, \citenamefont {Scherer}, \citenamefont
  {Sanders}, \citenamefont {Low}, \citenamefont {Wiebe}, \citenamefont
  {Gidney},\ and\ \citenamefont {Babbush}}]{Berry2018}%
  \BibitemOpen
  \bibfield  {author} {\bibinfo {author} {\bibfnamefont {Dominic~W}\
  \bibnamefont {Berry}}, \bibinfo {author} {\bibfnamefont {Maria}\ \bibnamefont
  {Kieferov{\'{a}}}}, \bibinfo {author} {\bibfnamefont {Artur}\ \bibnamefont
  {Scherer}}, \bibinfo {author} {\bibfnamefont {Yuval~R}\ \bibnamefont
  {Sanders}}, \bibinfo {author} {\bibfnamefont {Guang~Hao}\ \bibnamefont
  {Low}}, \bibinfo {author} {\bibfnamefont {Nathan}\ \bibnamefont {Wiebe}},
  \bibinfo {author} {\bibfnamefont {Craig}\ \bibnamefont {Gidney}}, \ and\
  \bibinfo {author} {\bibfnamefont {Ryan}\ \bibnamefont {Babbush}},\ }\bibfield
   {title} {\enquote {\bibinfo {title} {Improved techniques for preparing
  eigenstates of fermionic {H}amiltonians},}\ }\href {\doibase
  10.1038/s41534-018-0071-5} {\bibfield  {journal} {\bibinfo  {journal} {npj
  Quantum Information}\ }\textbf {\bibinfo {volume} {4}},\ \bibinfo {pages}
  {22} (\bibinfo {year} {2018})}\BibitemShut {NoStop}%
\bibitem [{\citenamefont {Boghosian}\ and\ \citenamefont
  {Taylor}(1998)}]{Boghosian1998}%
  \BibitemOpen
  \bibfield  {author} {\bibinfo {author} {\bibfnamefont {Bruce~M}\ \bibnamefont
  {Boghosian}}\ and\ \bibinfo {author} {\bibfnamefont {Washington}\
  \bibnamefont {Taylor}},\ }\bibfield  {title} {\enquote {\bibinfo {title}
  {{Simulating quantum mechanics on a quantum computer}},}\ }\href {\doibase
  10.1016/S0167-2789(98)00042-6} {\bibfield  {journal} {\bibinfo  {journal}
  {Physica D: Nonlinear Phenomena}\ }\textbf {\bibinfo {volume} {120}},\
  \bibinfo {pages} {30--42} (\bibinfo {year} {1998})}\BibitemShut {NoStop}%
\bibitem [{\citenamefont {Zalka}(1998)}]{Zalka1998}%
  \BibitemOpen
  \bibfield  {author} {\bibinfo {author} {\bibfnamefont {Christof}\
  \bibnamefont {Zalka}},\ }\bibfield  {title} {\enquote {\bibinfo {title}
  {Efficient simulation of quantum systems by quantum computers},}\ }\href
  {\doibase 10.1002/(SICI)1521-3978(199811)46:6/8<877::AID-PROP877>3.0.CO;2-A}
  {\bibfield  {journal} {\bibinfo  {journal} {Fortschritte der Physik}\
  }\textbf {\bibinfo {volume} {46}},\ \bibinfo {pages} {877--879} (\bibinfo
  {year} {1998})}\BibitemShut {NoStop}%
\bibitem [{\citenamefont {Lidar}\ and\ \citenamefont {Wang}(1999)}]{Lidar1999}%
  \BibitemOpen
  \bibfield  {author} {\bibinfo {author} {\bibfnamefont {Daniel~A}\
  \bibnamefont {Lidar}}\ and\ \bibinfo {author} {\bibfnamefont {Haobin}\
  \bibnamefont {Wang}},\ }\bibfield  {title} {\enquote {\bibinfo {title}
  {{Calculating the thermal rate constant with exponential speedup on a quantum
  computer}},}\ }\href {\doibase 10.1103/PhysRevE.59.2429} {\bibfield
  {journal} {\bibinfo  {journal} {Physical Review E}\ }\textbf {\bibinfo
  {volume} {59}},\ \bibinfo {pages} {2429--2438} (\bibinfo {year}
  {1999})}\BibitemShut {NoStop}%
\bibitem [{\citenamefont {Kassal}\ \emph {et~al.}(2008)\citenamefont {Kassal},
  \citenamefont {Jordan}, \citenamefont {Love}, \citenamefont {Mohseni},\ and\
  \citenamefont {Aspuru-Guzik}}]{Kassal2008}%
  \BibitemOpen
  \bibfield  {author} {\bibinfo {author} {\bibfnamefont {Ivan}\ \bibnamefont
  {Kassal}}, \bibinfo {author} {\bibfnamefont {Stephen~P}\ \bibnamefont
  {Jordan}}, \bibinfo {author} {\bibfnamefont {Peter~J}\ \bibnamefont {Love}},
  \bibinfo {author} {\bibfnamefont {Masoud}\ \bibnamefont {Mohseni}}, \ and\
  \bibinfo {author} {\bibfnamefont {Al{\'a}n}\ \bibnamefont {Aspuru-Guzik}},\
  }\bibfield  {title} {\enquote {\bibinfo {title} {{Polynomial-time quantum
  algorithm for the simulation of chemical dynamics}},}\ }\href
  {http://www.pnas.org/content/105/48/18681.abstract} {\bibfield  {journal}
  {\bibinfo  {journal} {Proceedings of the National Academy of Sciences}\
  }\textbf {\bibinfo {volume} {105}},\ \bibinfo {pages} {18681--18686}
  (\bibinfo {year} {2008})}\BibitemShut {NoStop}%
\bibitem [{\citenamefont {Ward}\ \emph {et~al.}(2008)\citenamefont {Ward},
  \citenamefont {Kassal},\ and\ \citenamefont {Aspuru-Guzik}}]{Ward2009}%
  \BibitemOpen
  \bibfield  {author} {\bibinfo {author} {\bibfnamefont {Nicholas~J}\
  \bibnamefont {Ward}}, \bibinfo {author} {\bibfnamefont {Ivan}\ \bibnamefont
  {Kassal}}, \ and\ \bibinfo {author} {\bibfnamefont {Al{\'a}n}\ \bibnamefont
  {Aspuru-Guzik}},\ }\bibfield  {title} {\enquote {\bibinfo {title}
  {{Preparation of many-body states for quantum simulation}},}\ }\href
  {\doibase http://dx.doi.org/10.1063/1.3115177} {\bibfield  {journal}
  {\bibinfo  {journal} {Journal Of Chemical Physics}\ }\textbf {\bibinfo
  {volume} {130}},\ \bibinfo {pages} {194105--194114} (\bibinfo {year}
  {2008})}\BibitemShut {NoStop}%
\bibitem [{\citenamefont {Kivlichan}\ \emph {et~al.}(2017)\citenamefont
  {Kivlichan}, \citenamefont {Wiebe}, \citenamefont {Babbush},\ and\
  \citenamefont {Aspuru-Guzik}}]{Kivlichan2016}%
  \BibitemOpen
  \bibfield  {author} {\bibinfo {author} {\bibfnamefont {Ian~D}\ \bibnamefont
  {Kivlichan}}, \bibinfo {author} {\bibfnamefont {Nathan}\ \bibnamefont
  {Wiebe}}, \bibinfo {author} {\bibfnamefont {Ryan}\ \bibnamefont {Babbush}}, \
  and\ \bibinfo {author} {\bibfnamefont {Al{\'a}n}\ \bibnamefont
  {Aspuru-Guzik}},\ }\bibfield  {title} {\enquote {\bibinfo {title} {{Bounding
  the costs of quantum simulation of many-body physics in real space}},}\
  }\href {http://iopscience.iop.org/article/10.1088/1751-8121/aa77b8}
  {\bibfield  {journal} {\bibinfo  {journal} {Journal of Physics A:
  Mathematical and Theoretical}\ }\textbf {\bibinfo {volume} {50}},\ \bibinfo
  {pages} {305301} (\bibinfo {year} {2017})}\BibitemShut {NoStop}%
\bibitem [{\citenamefont {Toloui}\ and\ \citenamefont
  {Love}(2013)}]{Toloui2013}%
  \BibitemOpen
  \bibfield  {author} {\bibinfo {author} {\bibfnamefont {Borzu}\ \bibnamefont
  {Toloui}}\ and\ \bibinfo {author} {\bibfnamefont {Peter~J}\ \bibnamefont
  {Love}},\ }\bibfield  {title} {\enquote {\bibinfo {title} {Quantum algorithms
  for quantum chemistry based on the sparsity of the {CI}-matrix},}\ }\href
  {http://arxiv.org/abs/1312.2579} {\bibfield  {journal} {\bibinfo  {journal}
  {arXiv:1312.2579}\ } (\bibinfo {year} {2013})}\BibitemShut {NoStop}%
\bibitem [{\citenamefont {Babbush}\ \emph
  {et~al.}(2018{\natexlab{c}})\citenamefont {Babbush}, \citenamefont {Berry},
  \citenamefont {Sanders}, \citenamefont {Kivlichan}, \citenamefont {Scherer},
  \citenamefont {Wei}, \citenamefont {Love},\ and\ \citenamefont
  {Aspuru-Guzik}}]{BabbushSparse2}%
  \BibitemOpen
  \bibfield  {author} {\bibinfo {author} {\bibfnamefont {Ryan}\ \bibnamefont
  {Babbush}}, \bibinfo {author} {\bibfnamefont {Dominic~W}\ \bibnamefont
  {Berry}}, \bibinfo {author} {\bibfnamefont {Yuval~R}\ \bibnamefont
  {Sanders}}, \bibinfo {author} {\bibfnamefont {Ian~D}\ \bibnamefont
  {Kivlichan}}, \bibinfo {author} {\bibfnamefont {Artur}\ \bibnamefont
  {Scherer}}, \bibinfo {author} {\bibfnamefont {Annie~Y}\ \bibnamefont {Wei}},
  \bibinfo {author} {\bibfnamefont {Peter~J}\ \bibnamefont {Love}}, \ and\
  \bibinfo {author} {\bibfnamefont {Al{\'a}n}\ \bibnamefont {Aspuru-Guzik}},\
  }\bibfield  {title} {\enquote {\bibinfo {title} {Exponentially more precise
  quantum simulation of fermions in the configuration interaction
  representation},}\ }\href
  {http://iopscience.iop.org/article/10.1088/2058-9565/aa9463/meta} {\bibfield
  {journal} {\bibinfo  {journal} {Quantum Science and Technology}\ }\textbf
  {\bibinfo {volume} {3}},\ \bibinfo {pages} {015006} (\bibinfo {year}
  {2018}{\natexlab{c}})}\BibitemShut {NoStop}%
\bibitem [{\citenamefont {Martin}(2004)}]{Martin2004}%
  \BibitemOpen
  \bibfield  {author} {\bibinfo {author} {\bibfnamefont {Richard}\ \bibnamefont
  {Martin}},\ }\href@noop {} {\emph {\bibinfo {title} {{Electronic
  Structure}}}}\ (\bibinfo  {publisher} {Cambridge University Press},\ \bibinfo
  {address} {Cambridge, U.K.},\ \bibinfo {year} {2004})\BibitemShut {NoStop}%
\bibitem [{\citenamefont {F{\"{u}}sti-Molnar}\ and\ \citenamefont
  {Pulay}(2002)}]{fusti2002accurate}%
  \BibitemOpen
  \bibfield  {author} {\bibinfo {author} {\bibfnamefont {László}\
  \bibnamefont {F{\"{u}}sti-Molnar}}\ and\ \bibinfo {author} {\bibfnamefont
  {Peter}\ \bibnamefont {Pulay}},\ }\bibfield  {title} {\enquote {\bibinfo
  {title} {{Accurate molecular integrals and energies using combined plane wave
  and gaussian basis sets in molecular electronic structure theory}},}\ }\href
  {http://aip.scitation.org/doi/abs/10.1063/1.1467901} {\bibfield  {journal}
  {\bibinfo  {journal} {The Journal of Chemical Physics}\ }\textbf {\bibinfo
  {volume} {116}},\ \bibinfo {pages} {7795--7805} (\bibinfo {year}
  {2002})}\BibitemShut {NoStop}%
\bibitem [{\citenamefont {Berry}\ \emph {et~al.}(2015)\citenamefont {Berry},
  \citenamefont {Childs}, \citenamefont {Cleve}, \citenamefont {Kothari},\ and\
  \citenamefont {Somma}}]{Berry2015}%
  \BibitemOpen
  \bibfield  {author} {\bibinfo {author} {\bibfnamefont {Dominic~W}\
  \bibnamefont {Berry}}, \bibinfo {author} {\bibfnamefont {Andrew~M}\
  \bibnamefont {Childs}}, \bibinfo {author} {\bibfnamefont {Richard}\
  \bibnamefont {Cleve}}, \bibinfo {author} {\bibfnamefont {Robin}\ \bibnamefont
  {Kothari}}, \ and\ \bibinfo {author} {\bibfnamefont {Rolando~D}\ \bibnamefont
  {Somma}},\ }\bibfield  {title} {\enquote {\bibinfo {title} {Simulating
  {H}amiltonian dynamics with a truncated {T}aylor series},}\ }\href {\doibase
  10.1103/PhysRevLett.114.090502} {\bibfield  {journal} {\bibinfo  {journal}
  {Physical Review Letters}\ }\textbf {\bibinfo {volume} {114}},\ \bibinfo
  {pages} {90502} (\bibinfo {year} {2015})}\BibitemShut {NoStop}%
\bibitem [{\citenamefont {Childs}\ and\ \citenamefont
  {Wiebe}(2012)}]{Childs2012}%
  \BibitemOpen
  \bibfield  {author} {\bibinfo {author} {\bibfnamefont {Andrew~M}\
  \bibnamefont {Childs}}\ and\ \bibinfo {author} {\bibfnamefont {Nathan}\
  \bibnamefont {Wiebe}},\ }\bibfield  {title} {\enquote {\bibinfo {title}
  {{Hamiltonian simulation using linear combinations of unitary operations}},}\
  }\href {https://dl.acm.org/citation.cfm?id=2481570} {\bibfield  {journal}
  {\bibinfo  {journal} {Quantum Information {\&} Computation}\ }\textbf
  {\bibinfo {volume} {12}},\ \bibinfo {pages} {901--924} (\bibinfo {year}
  {2012})}\BibitemShut {NoStop}%
\bibitem [{\citenamefont {Berry}\ \emph {et~al.}(2014)\citenamefont {Berry},
  \citenamefont {Childs}, \citenamefont {Cleve}, \citenamefont {Kothari},\ and\
  \citenamefont {Somma}}]{Berry2013}%
  \BibitemOpen
  \bibfield  {author} {\bibinfo {author} {\bibfnamefont {Dominic~W}\
  \bibnamefont {Berry}}, \bibinfo {author} {\bibfnamefont {Andrew~M}\
  \bibnamefont {Childs}}, \bibinfo {author} {\bibfnamefont {Richard}\
  \bibnamefont {Cleve}}, \bibinfo {author} {\bibfnamefont {Robin}\ \bibnamefont
  {Kothari}}, \ and\ \bibinfo {author} {\bibfnamefont {Rolando~D}\ \bibnamefont
  {Somma}},\ }\bibfield  {title} {\enquote {\bibinfo {title} {Exponential
  improvement in precision for simulating sparse {H}amiltonians},}\ }in\ \href
  {https://doi.org/10.1145/2591796.2591854} {\emph {\bibinfo {booktitle} {STOC
  '14 Proceedings of the 46th Annual ACM Symposium on Theory of Computing}}}\
  (\bibinfo {year} {2014})\ pp.\ \bibinfo {pages} {283--292}\BibitemShut
  {NoStop}%
\bibitem [{\citenamefont {Low}\ \emph {et~al.}(2018)\citenamefont {Low},
  \citenamefont {Kliuchnikov},\ and\ \citenamefont {Schaeffer}}]{Low2018a}%
  \BibitemOpen
  \bibfield  {author} {\bibinfo {author} {\bibfnamefont {Guang~Hao}\
  \bibnamefont {Low}}, \bibinfo {author} {\bibfnamefont {Vadym}\ \bibnamefont
  {Kliuchnikov}}, \ and\ \bibinfo {author} {\bibfnamefont {Luke}\ \bibnamefont
  {Schaeffer}},\ }\bibfield  {title} {\enquote {\bibinfo {title} {{Trading
  T-gates for dirty qubits in state preparation and unitary synthesis}},}\
  }\href {http://arxiv.org/abs/1812.00954} {\bibfield  {journal} {\bibinfo
  {journal} {arXiv:1812.00954}\ } (\bibinfo {year} {2018})}\BibitemShut
  {NoStop}%
\bibitem [{\citenamefont {Bailey}\ \emph {et~al.}(2010)\citenamefont {Bailey},
  \citenamefont {Borwein},\ and\ \citenamefont {Crandall}}]{Bailey2010}%
  \BibitemOpen
  \bibfield  {author} {\bibinfo {author} {\bibfnamefont {D.~H.}\ \bibnamefont
  {Bailey}}, \bibinfo {author} {\bibfnamefont {J.~M.}\ \bibnamefont {Borwein}},
  \ and\ \bibinfo {author} {\bibfnamefont {R.~E.}\ \bibnamefont {Crandall}},\
  }\bibfield  {title} {\enquote {\bibinfo {title} {{Advances in the theory of
  box integrals}},}\ }\href {\doibase 10.1090/S0025-5718-10-02338-0} {\bibfield
   {journal} {\bibinfo  {journal} {Mathematics of Computation}\ }\textbf
  {\bibinfo {volume} {79}},\ \bibinfo {pages} {1839--1866} (\bibinfo {year}
  {2010})}\BibitemShut {NoStop}%
\bibitem [{\citenamefont {Reiher}\ \emph {et~al.}(2017)\citenamefont {Reiher},
  \citenamefont {Wiebe}, \citenamefont {Svore}, \citenamefont {Wecker},\ and\
  \citenamefont {Troyer}}]{Reiher2017}%
  \BibitemOpen
  \bibfield  {author} {\bibinfo {author} {\bibfnamefont {Markus}\ \bibnamefont
  {Reiher}}, \bibinfo {author} {\bibfnamefont {Nathan}\ \bibnamefont {Wiebe}},
  \bibinfo {author} {\bibfnamefont {Krysta~M}\ \bibnamefont {Svore}}, \bibinfo
  {author} {\bibfnamefont {Dave}\ \bibnamefont {Wecker}}, \ and\ \bibinfo
  {author} {\bibfnamefont {Matthias}\ \bibnamefont {Troyer}},\ }\bibfield
  {title} {\enquote {\bibinfo {title} {Elucidating reaction mechanisms on
  quantum computers},}\ }\href
  {http://www.pnas.org/content/114/29/7555.abstract} {\bibfield  {journal}
  {\bibinfo  {journal} {Proceedings of the National Academy of Sciences}\
  }\textbf {\bibinfo {volume} {114}},\ \bibinfo {pages} {7555--7560} (\bibinfo
  {year} {2017})}\BibitemShut {NoStop}%
\bibitem [{\citenamefont {Motta}\ \emph {et~al.}(2018)\citenamefont {Motta},
  \citenamefont {Ye}, \citenamefont {McClean}, \citenamefont {Li},
  \citenamefont {Minnich}, \citenamefont {Babbush},\ and\ \citenamefont
  {Chan}}]{Motta2018}%
  \BibitemOpen
  \bibfield  {author} {\bibinfo {author} {\bibfnamefont {Mario}\ \bibnamefont
  {Motta}}, \bibinfo {author} {\bibfnamefont {Erika}\ \bibnamefont {Ye}},
  \bibinfo {author} {\bibfnamefont {Jarrod~R.}\ \bibnamefont {McClean}},
  \bibinfo {author} {\bibfnamefont {Zhendong}\ \bibnamefont {Li}}, \bibinfo
  {author} {\bibfnamefont {Austin~J.}\ \bibnamefont {Minnich}}, \bibinfo
  {author} {\bibfnamefont {Ryan}\ \bibnamefont {Babbush}}, \ and\ \bibinfo
  {author} {\bibfnamefont {Garnet Kin-Lic}\ \bibnamefont {Chan}},\ }\bibfield
  {title} {\enquote {\bibinfo {title} {Low rank representations for quantum
  simulation of electronic structure},}\ }\href
  {http://arxiv.org/abs/1808.02625} {\bibfield  {journal} {\bibinfo  {journal}
  {arXiv:1808.02625}\ } (\bibinfo {year} {2018})}\BibitemShut {NoStop}%
\bibitem [{\citenamefont {Li}\ \emph {et~al.}(2019)\citenamefont {Li},
  \citenamefont {Li}, \citenamefont {Dattani}, \citenamefont {Umrigar},\ and\
  \citenamefont {Chan}}]{Zhendong2018}%
  \BibitemOpen
  \bibfield  {author} {\bibinfo {author} {\bibfnamefont {Zhendong}\
  \bibnamefont {Li}}, \bibinfo {author} {\bibfnamefont {Junhao}\ \bibnamefont
  {Li}}, \bibinfo {author} {\bibfnamefont {Nikesh~S.}\ \bibnamefont {Dattani}},
  \bibinfo {author} {\bibfnamefont {C.~J.}\ \bibnamefont {Umrigar}}, \ and\
  \bibinfo {author} {\bibfnamefont {Garnet Kin-Lic}\ \bibnamefont {Chan}},\
  }\bibfield  {title} {\enquote {\bibinfo {title} {The electronic complexity of
  the ground-state of the {FeMo} cofactor of nitrogenase as relevant to quantum
  simulations},}\ }\href {https://doi.org/10.1063/1.5063376} {\bibfield
  {journal} {\bibinfo  {journal} {The Journal of Chemical Physics}\ }\textbf
  {\bibinfo {volume} {150}},\ \bibinfo {pages} {024302} (\bibinfo {year}
  {2019})}\BibitemShut {NoStop}%
\bibitem [{\citenamefont {Takeshita}\ \emph {et~al.}(2019)\citenamefont
  {Takeshita}, \citenamefont {Rubin}, \citenamefont {Jiang}, \citenamefont
  {Lee}, \citenamefont {Babbush},\ and\ \citenamefont
  {McClean}}]{Takeshita2019}%
  \BibitemOpen
  \bibfield  {author} {\bibinfo {author} {\bibfnamefont {Tyler}\ \bibnamefont
  {Takeshita}}, \bibinfo {author} {\bibfnamefont {Nicholas~C.}\ \bibnamefont
  {Rubin}}, \bibinfo {author} {\bibfnamefont {Zhang}\ \bibnamefont {Jiang}},
  \bibinfo {author} {\bibfnamefont {Eunseok}\ \bibnamefont {Lee}}, \bibinfo
  {author} {\bibfnamefont {Ryan}\ \bibnamefont {Babbush}}, \ and\ \bibinfo
  {author} {\bibfnamefont {Jarrod~R.}\ \bibnamefont {McClean}},\ }\bibfield
  {title} {\enquote {\bibinfo {title} {{Increasing the representation accuracy
  of quantum simulations of chemistry without extra quantum resources}},}\
  }\href {https://arxiv.org/abs/1902.10679} {\bibfield  {journal} {\bibinfo
  {journal} {arXiv:1902.10679}\ } (\bibinfo {year} {2019})}\BibitemShut
  {NoStop}%
\bibitem [{\citenamefont {Low}\ and\ \citenamefont {Chuang}(2017)}]{Low2017}%
  \BibitemOpen
  \bibfield  {author} {\bibinfo {author} {\bibfnamefont {Guang~Hao}\
  \bibnamefont {Low}}\ and\ \bibinfo {author} {\bibfnamefont {Isaac~L}\
  \bibnamefont {Chuang}},\ }\bibfield  {title} {\enquote {\bibinfo {title}
  {Optimal {H}amiltonian simulation by quantum signal processing},}\ }\href
  {\doibase 10.1103/PhysRevLett.118.010501} {\bibfield  {journal} {\bibinfo
  {journal} {Physical Review Letters}\ }\textbf {\bibinfo {volume} {118}},\
  \bibinfo {pages} {010501} (\bibinfo {year} {2017})}\BibitemShut {NoStop}%
\end{thebibliography}%

\end{document}